\documentclass[10pt,twocolumn]{article}

\usepackage[letterpaper,margin=0.75in]{geometry}
\usepackage{amsmath,amssymb,amsfonts}
\usepackage{graphicx}
\usepackage{textcomp}
\usepackage{xcolor}
\usepackage{booktabs}
\usepackage{url}
\usepackage[numbers,sort&compress]{natbib}
\usepackage{bm}

\providecommand{\SI}[2]{#1\,#2}
\providecommand{\si}[1]{#1}
\providecommand{\num}[1]{#1}
\providecommand{\meter}{m}
\providecommand{\second}{s}
\providecommand{\radian}{rad}
\providecommand{\kilogram}{kg}
\providecommand{\newton}{N}

\providecommand{\milli}{m}
\providecommand{\per}{/}
\usepackage{titlesec}
\usepackage{authblk}
\usepackage{abstract}

\titleformat{\section}{\normalfont\large\bfseries}{\thesection.}{0.5em}{\MakeUppercase}
\titleformat{\subsection}{\normalfont\normalsize\bfseries\itshape}{\thesubsection)}{0.5em}{}

\graphicspath{{figures/}}

\newcommand{\ey}{e_y}
\newcommand{\epsiang}{e_{\psi}}
\newcommand{\dt}{\Delta t}

\title{A Hybrid End-to-End and Modular Control Architecture
Toward Safe Vehicle Lateral Control: \\
Combining Soft Actor-Critic with Model Predictive Control}

\author{Farzaneh Tatari\\
\small Independent Researcher, Novi, MI, USA\\
\small Email: \texttt{fa\_tatari@yahoo.com} \quad
ORCID: \texttt{0000-0001-5176-3372}\\
\scriptsize\itshape This work was performed independently by the
author on personal time and does not reflect the views, positions,
or products of any employer.}
\date{}

\begin{document}

\twocolumn[
\begin{@twocolumnfalse}
\maketitle
\begin{abstract}\noindent
Connected and automated vehicles demand lateral controllers that are
simultaneously accurate, low-effort, and safe under model error and
sensor noise. Modular controllers such as model predictive control
(MPC) are interpretable and constraint-aware but rely on accurate
models and hand-tuned weights. End-to-end learned policies, in
particular continuous-action deep reinforcement learning, are
adaptable and require no hand-designed control law, but offer no
intrinsic safety guarantees and limited interpretability. This
paper presents a hybrid architecture that combines an end-to-end
Soft Actor-Critic (SAC) policy with a constrained linear MPC into a
single steering command, using the MPC's first-step optimum as the
model-based anchor and a single monotone blending coefficient that
interpolates between the two paradigms. The architecture is
evaluated on a linearized lateral bicycle model against a PID
baseline, a tuned linear MPC, and a stand-alone SAC policy, across
nominal, single-axis robustness, and multi-initial-condition
ensemble experiments. The hybrid retains the tracking quality of
stand-alone SAC while remaining inside the MPC's actuator envelope
and preserving a deterministic, model-based contribution to every
steering command. The architecture provides an actuator-envelope
guarantee by construction but does \emph{not} establish recursive
feasibility or terminal invariance, and the closed-form blend does
not prevent all corner-case divergences at the boundary of the
training distribution. A corner-case analysis shows that the blend
attenuates but cannot prevent failure under distribution shift,
motivating a connectivity-aware extension in which the blending
coefficient is scheduled by vehicle-to-everything (V2X) signals to
restore model-based authority. Limitations and a path toward a
constrained-QP predictive safety filter are discussed.
\end{abstract}
\medskip
\noindent\textbf{Keywords:} \textit{Autonomous vehicles, lateral control,
model predictive control, soft actor-critic, safe reinforcement
learning, hybrid control, end-to-end learning.}
\bigskip
\end{@twocolumnfalse}
]

\section{Introduction}
Connected and automated vehicles (CAVs) require lateral controllers that
are simultaneously accurate, low-effort, and safe under modeling error,
sensor noise, and unmodeled disturbances. Two broad families of
controllers dominate the literature. \emph{Modular} approaches, including PID,
gain-scheduled feedback, and model predictive control (MPC), rely on a
physics-based vehicle model and explicit constraints, so they are
interpretable and easy to certify, but they require careful tuning and
their performance degrades when the underlying model is inaccurate
\cite{rajamani2011vehicle,kong2015kinematic,funke2017collision}.
\emph{End-to-end} approaches train neural-network policies via
either imitation learning from driving demonstrations or deep
reinforcement learning (RL) on a tracking reward
\cite{kendall2019learn,bojarski2016end,haarnoja2018soft}. End-to-end
policies can adapt to nonlinearities and unmodeled dynamics but offer
weak guarantees on input feasibility, comfort, and safety, and they
are harder to debug and tune.

Hybrid approaches combine end-to-end components with modular or
hierarchical components. Such hybrids
attempt to retain the data-driven adaptability of learned policies
while preserving the safety, interpretability, and predictability of
modular control. Promising directions include control-barrier-function
shields \cite{cheng2019end}, predictive safety filters
\cite{wabersich2018linear,wabersich2021predictive}, and
learning-augmented MPC \cite{kabzan2019learning,hewing2020learning}.

This paper develops and evaluates one concrete instance of such a
hybrid for vehicle \emph{lateral} control. The contributions are as
follows.
\begin{enumerate}
\item Four lateral controllers, namely PID, linear MPC, Soft
Actor-Critic (SAC), and a hybrid, are benchmarked on a four-state
linearized lateral bicycle model at a constant longitudinal speed
of \SI{15}{\meter\per\second}, using shared metrics (RMSE, peak
error, mean absolute steering, final error).
\item The linear MPC is formulated with input and input-rate
constraints, implemented as a quadratic program (QP) over a horizon
of \num{15} steps, and tuned via grid search over \num{243} weight
configurations.
\item A SAC policy is trained from scratch on the same plant in an
episodic simulator with randomized initial conditions, using a
quadratic reward over lateral-error, heading-error, steering
magnitude, and steering change-rate, and is evaluated
deterministically.
\item A hybrid architecture is proposed in which the MPC's
first-step optimum and the SAC policy are linearly combined through
a single blending coefficient, with the resulting steering command
saturated to the same actuator limits used by the MPC.
\item A robustness study is conducted covering large initial errors,
process noise, and tire-stiffness mismatch.
\item A detailed discussion of the limitations of the current hybrid
implementation is provided, together with a concrete path toward
the predictive safety filter formulated in
Section~\ref{sec:hybrid}.
\end{enumerate}

The remainder of the paper is organized as follows.
Section~\ref{sec:related} reviews related work as modular, end-to-end,
and hybrid families. Section~\ref{sec:dynamics} introduces the
vehicle model. Section~\ref{sec:mpc} presents the modular component
(linear MPC); Section~\ref{sec:sac} presents the end-to-end component
(SAC); Section~\ref{sec:hybrid} presents the hybrid that combines
them. Section~\ref{sec:experiments} describes the experimental setup,
Section~\ref{sec:results} presents nominal, tuning, robustness, and
multi-IC ensemble results, and Section~\ref{sec:discussion} closes
with what each side contributes, limitations, and a
connectivity-aware extension path.
Section~\ref{sec:conclusion} concludes.

\section{Related Work}\label{sec:related}

The related work is organized as follows: modular controllers,
end-to-end learned policies, and hybrid architectures that combine
the two.

\subsection{Modular Approaches for Vehicle Lateral Control}
Modular controllers rely on a physics-based vehicle model and
hand-designed control law. PID and gain-scheduled feedback have a long
history in production vehicles, but in the research literature MPC has
become the dominant choice for CAV lateral control because it handles
input and state constraints explicitly and exploits a predictive
vehicle model
\cite{falcone2007predictive,borrelli2005mpc,funke2017collision,rajamani2011vehicle}.
Linearized lateral bicycle models are widely used because they yield
QPs that can be solved in real time
\cite{kong2015kinematic,stellato2020osqp}. Tube and stochastic MPC
formulations
\cite{kohler2020computationally,hewing2020learning}
extend the framework to handle bounded disturbances and parameter
uncertainty (tube MPC tightens constraints by a robust invariant set
to enforce worst-case feasibility, while stochastic MPC replaces hard
constraints with chance constraints under an assumed disturbance
distribution). The strengths of the modular family are interpretability,
explicit constraint handling, and amenability to safety analysis; its
main weakness is sensitivity to model fidelity and tuning effort.

\subsection{End-to-End Learned Approaches}
End-to-end approaches replace the hand-designed control law with a
neural-network policy trained directly on a tracking or driving
reward. Imitation learning from human-driven trajectories has been
demonstrated at the full vehicle scale \cite{bojarski2016end}.
Continuous-action deep RL algorithms, including Deep Deterministic Policy Gradient (DDPG) \cite{lillicrap2015continuous},
Twin Delayed DDPG (TD3) \cite{fujimoto2018addressing}, and Soft Actor-Critic (SAC)
\cite{haarnoja2018soft,haarnoja2018sacapps}, are now standard for
steering tasks because their policies are differentiable, sample
efficient, and easy to deploy at the actuator rate. SAC's
maximum-entropy objective in particular yields stochastic exploration
that is robust to hyperparameter choice, in particular the entropy
temperature, which SAC tunes automatically
\cite{haarnoja2018sacapps}. This family of methods has been
demonstrated end-to-end on real vehicles \cite{kendall2019learn}
and surveyed extensively for autonomous driving
\cite{kiran2021deep}. The strengths of the end-to-end
family are adaptability to nonlinearity and unmodeled dynamics, and
the absence of a hand-tuned control law; its main weaknesses are the
lack of intrinsic safety guarantees, weak interpretability, and
brittleness under distribution shift (states or plant parameters
outside the training distribution).

\subsection{Hybrid Architectures Combining End-to-End with Modular Components}
The recent literature increasingly combines a learned end-to-end
component with a modular safety or tracking component, in an explicit
attempt to retain the strengths of both families. Predictive safety
filters \cite{wabersich2018linear,wabersich2021predictive,gros2020safe}
treat the learned policy's output as a soft preference and project it
onto the closest input that the modular MPC can certify as safe.
Control-barrier-function (CBF) shields enforce forward invariance of a
safe set on top of a learned controller \cite{cheng2019end}.
Learning-augmented MPC equips a model-based controller with a learned
residual or value function
\cite{hewing2020learning,kabzan2019learning}. At the architectural
level, several works combine iteratively-learned terminal sets
\cite{rosolia2017learning} or learned subgoal recommendation
\cite{brito2021where} with a downstream MPC. The architecture we present in this paper falls into the
predictive-safety-filter family. The implementation evaluated here is
a simplified version: instead of re-solving the full constrained QP
with the SAC action as a soft preference, we blend the MPC's
first-step optimum with the SAC action in closed form.

\section{Vehicle Dynamics Model}\label{sec:dynamics}
We use a linearized lateral bicycle model with state
\begin{equation}
x = \begin{bmatrix} \ey & \epsiang & v_y & r \end{bmatrix}^{\!\top},
\end{equation}
where $\ey$ is the lateral path-following error, $\epsiang$ is the
heading error relative to the path, $v_y$ is the body-frame lateral
velocity, and $r$ is the yaw rate. The control input is the front
steering angle $\delta$. The longitudinal speed $v_x$ is held constant
at the nominal value (\SI{15}{\meter\per\second}). The continuous-time
dynamics are
\begin{align}
\dot{\ey} &= v_y + v_x\,\epsiang, \\
\dot{\epsiang} &= r, \\
\dot{v}_y &= -\frac{2C_f + 2C_r}{m\,v_x} v_y
            - \!\left( v_x + \frac{2C_f l_f - 2C_r l_r}{m\,v_x} \right)\!r
            + \frac{2C_f}{m}\delta, \\
\dot{r}   &= -\frac{2C_f l_f - 2C_r l_r}{I_z\,v_x} v_y
            - \frac{2C_f l_f^2 + 2C_r l_r^2}{I_z\,v_x} r
            + \frac{2C_f l_f}{I_z}\delta.
\end{align}
The discrete-time model used by all controllers and the simulator is
the forward-Euler discretization of these equations with sample time
$\dt = \SI{0.05}{\second}$,
\begin{equation}
x_{k+1} = A\,x_k + B\,\delta_k.
\label{eq:linmodel}
\end{equation}
Here $A \in \mathbb{R}^{4\times 4}$ is the discrete-time state matrix
and $B \in \mathbb{R}^{4\times 1}$ is the discrete-time input matrix,
obtained by forward-Euler discretization of the continuous-time
dynamics above: $A = I + \dt \, A_c$ and $B = \dt \, B_c$, where the
continuous-time state and input Jacobians, evaluated at the nominal
$v_x$, are
{\setlength{\arraycolsep}{3pt}\small
\begin{equation}
A_c = \frac{\partial f}{\partial x} =
\begin{bmatrix}
0 & v_x & 1 & 0 \\
0 & 0 & 0 & 1 \\
0 & 0 & -\dfrac{2C_f + 2C_r}{m\,v_x}
       & -v_x - \dfrac{2C_f l_f - 2C_r l_r}{m\,v_x} \\[4pt]
0 & 0 & -\dfrac{2C_f l_f - 2C_r l_r}{I_z\,v_x}
       & -\dfrac{2C_f l_f^{2} + 2C_r l_r^{2}}{I_z\,v_x}
\end{bmatrix},
\end{equation}}
\begin{equation}
B_c = \frac{\partial f}{\partial \delta} =
\begin{bmatrix}
0 \\ 0 \\ \dfrac{2C_f}{m} \\[4pt] \dfrac{2C_f l_f}{I_z}
\end{bmatrix},
\end{equation}
where $f(x, \delta)$ denotes the right-hand side of the
continuous-time dynamics above.
The vehicle parameters are summarized in Table~\ref{tab:vehicle}.

\begin{table}[t]
\centering
\caption{Nominal vehicle parameters.}
\label{tab:vehicle}
\begin{tabular}{@{}lll@{}}
\toprule
Symbol & Value & Description \\ \midrule
$m$        & \SI{1600}{\kilogram}             & Mass \\
$I_z$      & 2500\,kg$\cdot$m$^2$              & Yaw inertia \\
$l_f$      & \SI{1.2}{\meter}                  & Front axle distance \\
$l_r$      & \SI{1.6}{\meter}                  & Rear axle distance \\
$C_f$      & \SI{80000}{\newton\per\radian}    & Front cornering stiffness \\
$C_r$      & \SI{80000}{\newton\per\radian}    & Rear cornering stiffness \\
$v_x$      & \SI{15}{\meter\per\second}        & Longitudinal speed \\
$\dt$      & \SI{0.05}{\second}                & Sample time \\
\bottomrule
\end{tabular}
\end{table}

\section{Modular Component: Linear MPC}\label{sec:mpc}
The linear MPC controller solves, at each step $k$, a finite-horizon
quadratic program over the input sequence
$U = [\delta_0,\dots,\delta_{N-1}]^{\!\top}$ (with $N=15$) given the
current state $x_0$,
\begin{align}
\min_{U}\; & \sum_{i=0}^{N-1}\!
\left( x_i^{\!\top} Q\, x_i + R\,\delta_i^{2} +
       R_d\,(\delta_i-\delta_{i-1})^{2} \right) \nonumber \\
&\quad + x_N^{\!\top} Q_f\, x_N \label{eq:mpccost}\\[2pt]
\text{s.t.}\; & X = \Phi\, x_0 + \Gamma\, U, \\
& |\delta_i| \le \delta_{\max} = \SI{0.4}{\radian}, \\
& |\delta_i - \delta_{i-1}| \le \delta_{\text{rate,max}} =
\SI{0.15}{\radian} \nonumber\\
&\quad\text{(i.e., } \SI{3}{\radian\per\second}\text{ at }
\dt = \SI{0.05}{\second}\text{)}, \\
& \delta_{-1} = \delta_{k-1}^{\star},
\end{align}
where $X = [x_1^{\!\top},\dots,x_N^{\!\top}]^{\!\top}$ is the
predicted state sequence and
\begin{equation}
\Phi =
\begin{bmatrix}
A \\ A^{2} \\ \vdots \\ A^{N}
\end{bmatrix},
\qquad
\Gamma =
\begin{bmatrix}
B            & 0            & \cdots & 0      \\
A B          & B            & \cdots & 0      \\
\vdots       & \vdots       & \ddots & \vdots \\
A^{N-1} B    & A^{N-2} B    & \cdots & B
\end{bmatrix},
\label{eq:phi-gamma}
\end{equation}
so that $X$ is fully determined by $x_0$ and $U$. Here
$\delta_{k-1}^{\star}$ is the previously applied steering command. The resulting QP in $U$ alone is solved
with the operator-splitting QP solver OSQP
\cite{stellato2020osqp}; the first element
$\delta_0^{\star}$ is applied to the plant and the remainder is
discarded.

\paragraph{Tuning.}
The state-weight diagonal entries $Q[0,0], Q[1,1], Q[3,3]$ (penalizing
lateral error, heading error, and yaw rate respectively), the input
weight $R$, and the input-rate weight $R_d$ were grid-searched over
three values each ($3^{5} = 243$ configurations) on the nominal
initial-condition recovery task; $Q[2,2]$ (lateral velocity) was held
fixed at $1.0$ and the terminal weight was tied as $Q_f = Q$. The
selected weights are
\begin{equation}
Q = Q_f = \mathrm{diag}(14, 5, 1, 2),\;
R = 0.06,\;
R_d = 1.5.
\end{equation}

\section{End-to-End Component: SAC Policy}\label{sec:sac}
A SAC policy \cite{haarnoja2018soft,haarnoja2018sacapps}
$\pi_\theta:\mathcal{S}\!\to\!\mathcal{A}$, with state space
$\mathcal{S} = \mathbb{R}^{4}$ and action space
$\mathcal{A} = [-0.4, 0.4]\,\si{\radian}$, is trained on the same
discretized lateral bicycle plant. The observation is the full state
$s_k = x_k = (\ey, \epsiang, v_y, r) \in \mathcal{S}$, and the action
is the steering command $a_k = \delta_k \in \mathcal{A}$. The reward
balances tracking accuracy, control effort, and steering smoothness,
\begin{equation}
r_k = -\,2.0\,\ey^2 - 1.0\,\epsiang^2
      - 0.05\,\delta_k^2 - 0.1\,(\delta_k - \delta_{k-1})^2.
\label{eq:reward}
\end{equation}
Each episode is initialized at $x_0 = (\ey^{(0)}, \epsiang^{(0)}, 0, 0)$
with $\ey^{(0)} \sim \mathcal{U}[-0.4, 0.4]$ and
$\epsiang^{(0)} \sim \mathcal{U}[-0.10, 0.10]$ (i.e., drawn
uniformly at random from these intervals at the start of each
episode), matching the initial-condition distribution used at
evaluation. Each episode runs for at most
\num{400} steps and terminates early if
$|\ey| > \SI{2.0}{\meter}$. We use the SAC implementation from
\texttt{stable-baselines3} \cite{raffin2021sb3}, an open-source
PyTorch library of reference RL algorithms. The actor and critic
networks use the library's default \texttt{MlpPolicy}: a feedforward
multi-layer perceptron with two hidden layers of \num{256} units and
ReLU activations. Training uses a replay buffer of capacity
\num{50000} transitions, \num{1000} learning-start transitions
(during which the agent only collects experience without gradient
updates), and a total of \num{150000} training steps (approximately
$400$-$600$ episodes per seed, depending on early-termination
frequency). The replay buffer operates as a first-in, first-out
sliding window: each environment step appends the latest transition,
and once the buffer reaches capacity each new transition evicts the
oldest. Each gradient update samples a minibatch of $256$ transitions
uniformly at random from the current buffer contents to update the
actor and critic networks; the buffer therefore stores past
experience for off-policy learning rather than feeding any single
update. The full training is repeated under five independent
seeds (\num{0}-\num{4}); all reported SAC and hybrid metrics are
aggregated over the five resulting checkpoints.

\section{Hybrid End-to-End / Modular Controller}\label{sec:hybrid}
The hybrid controller blends two signals: the trained SAC steering
$\delta_{\mathrm{SAC}}$ and the linear MPC's first-step optimum
$\delta_{\mathrm{MPC}}$ at the current state. Concretely, at each step
$k$ the MPC of Section~\ref{sec:mpc} is solved on the same state $x_k$
that SAC observes, and the first input of its solution is taken as
$\delta_{\mathrm{MPC}} = \delta_0^{\star}(x_k)$. With a single blending
coefficient $\lambda \ge 0$, define
\begin{equation}
\alpha(\lambda) = \frac{\lambda}{1+\lambda} \in [0,1).
\end{equation}
The applied command is the saturated convex combination
\begin{equation}
\delta_k = \mathrm{sat}_{\delta_{\max}}\!\left(
(1-\alpha(\lambda))\,\delta_{\mathrm{MPC}} +
\alpha(\lambda)\,\delta_{\mathrm{SAC}}
\right),
\label{eq:hybridcmd}
\end{equation}
where $\mathrm{sat}_{\delta_{\max}}(u) = \max\!\left(-\delta_{\max},\,
\min(u, \delta_{\max})\right)$ clips its argument to the actuator
interval $[-\delta_{\max}, +\delta_{\max}]$ with
$\delta_{\max} = \SI{0.4}{\radian}$, i.e.\ the same hard actuator
bound enforced by the MPC in Section~\ref{sec:mpc}.
Equation~\eqref{eq:hybridcmd} interpolates between the model-based
controller ($\lambda \to 0$, $\alpha \to 0$, recovering the pure MPC
of Section~\ref{sec:mpc}) and the end-to-end SAC policy
($\lambda \to \infty$, $\alpha \to 1$). The form
$\alpha = \lambda/(1+\lambda)$ is a smooth, monotone parameterization
of this trade-off and is invariant to the units of $\lambda$.

\paragraph{Relation to a true predictive safety filter.}
A true predictive safety filter \cite{wabersich2018linear,
wabersich2021predictive} would solve, at each step, a constrained QP
\begin{align}
\min_{U}\; & \sum_{i=0}^{N-1}\! x_i^{\!\top} Q\, x_i +
\mu\,(\delta_0 - \delta_{\mathrm{SAC}})^{2} \nonumber\\
&\quad + R\,\delta_i^{2} + R_d\,(\delta_i-\delta_{i-1})^{2}
\label{eq:safetyfilter}\\
\text{s.t.}\; & X = \Phi\, x_0 + \Gamma\, U,\;
|\delta_i|\le\delta_{\max},\;
|\delta_i-\delta_{i-1}|\le\delta_{\text{rate,max}}, \nonumber
\end{align}
where the SAC action enters as a soft preference with weight
$\mu \ge 0$ on the deviation from $\delta_{\mathrm{SAC}}$. The
implementation evaluated in this paper does \emph{not} solve
\eqref{eq:safetyfilter}. At each
step we instead use the MPC's first-step optimum
$\delta_0^{\star}$ from the standard MPC of Section~\ref{sec:mpc}
as the model-based anchor, and apply the analytic blend
\eqref{eq:hybridcmd}.

\paragraph{What the hybrid does and does not guarantee.}
\emph{Guaranteed by construction:} the per-step actuator magnitude
bound $|\delta_k| \le \delta_{\max}$ via the saturation
$\mathrm{sat}_{\delta_{\max}}(\cdot)$, and a non-zero, deterministic,
model-based contribution to every steering command with magnitude at
most $(1-\alpha)\,\delta_{\max}$.
\emph{Not guaranteed:} the input-rate constraint over the blended
horizon (the rate bound is enforced only inside $\delta_{\mathrm{MPC}}$'s
own prediction, not on the blended output); recursive feasibility;
terminal invariance; and, at the chosen $\alpha \approx 0.92$,
prevention of corner-case divergences when $\delta_{\mathrm{SAC}}$
saturates in the wrong direction (see Section~\ref{subsec:multic}).
These are exactly the properties that the full constrained-QP
predictive safety filter \eqref{eq:safetyfilter} is designed to
provide and that we identify as the direct next step for the
architecture.

\section{Experimental Setup}\label{sec:experiments}
All controllers are evaluated on the same discretized plant
\eqref{eq:linmodel} with the parameters of Table~\ref{tab:vehicle}.
Unless stated otherwise, the nominal episode initializes at
$x_0 = (0.2,\, 0.05,\, 0,\, 0)$ and runs for \num{400} steps
($T = \SI{20}{\second}$). For each rollout (a single closed-loop
simulation from $x_0$ to the final step) we report the
root-mean-square lateral error
\begin{equation}
\mathrm{RMSE}(\ey) = \sqrt{\tfrac{1}{T}\textstyle\sum_k \ey^2(k)},
\end{equation}
the peak absolute lateral error $\max_k|\ey(k)|$, the mean absolute
steering effort $\overline{|\delta|}$, and the final lateral error
$\ey(T)$.

\subsection{Hybrid Tuning}
The blending parameter $\lambda$ is selected on the nominal task by
sweeping $\lambda \in \{1,2,3,5,8,10,12,15\}$ and minimizing the
composite score
\begin{equation}
J = \mathrm{RMSE}(\ey) + 0.5\,\max_k|\ey(k)| +
0.5\,\overline{|\delta|}.
\label{eq:lamscore}
\end{equation}

\subsection{Robustness Cases}
Four perturbation scenarios test sensitivity to off-nominal conditions:
\begin{description}
\item[Case A:] Large initial lateral error, $x_0 = (0.4,\,0.05,\,0,\,0)$.
\item[Case B:] Large initial heading error, $x_0 = (0.2,\,0.10,\,0,\,0)$.
\item[Case C:] Zero-mean Gaussian process noise
$w_k \sim \mathcal{N}(0, \mathrm{diag}(\sigma_x^2))$ added to the
state at every step, with one standard deviation per state component,
$\sigma_{\ey} = 10^{-3}\,\si{\meter}$,
$\sigma_{\epsiang} = 5\!\times\!10^{-4}\,\si{\radian}$,
$\sigma_{v_y} = 5\!\times\!10^{-3}\,\si{\meter\per\second}$,
$\sigma_{r} = 2\!\times\!10^{-3}\,\si{\radian\per\second}$.
\item[Case D:] Tire-stiffness mismatch, controllers and SAC trained
with $C_f, C_r = \SI{80000}{\newton\per\radian}$ but the simulator uses
$0.8\,C_f$ and $0.8\,C_r$.
\end{description}
\subsection{Multi-Seed and Multi-IC Aggregation}
All reported SAC and hybrid metrics aggregate over the five
independent SAC training seeds (\num{0}-\num{4}) of
Section~\ref{sec:sac}; PID and MPC do not depend on the SAC seed and
their numbers are deterministic. In
Sections~\ref{subsec:nominal} and~\ref{subsec:robustness} (the
nominal task and the four robustness cases), SAC and hybrid metrics
are reported as mean $\pm$ standard deviation over the five seeds. The multi-IC ensemble
(Section~\ref{subsec:multic}) additionally aggregates over a
$5\!\times\!5$ initial-condition grid,
$\ey \in \{-0.4,-0.2,0,0.2,0.4\}\,\si{\meter}$ and
$\epsiang \in \{-0.10,-0.05,0,0.05,0.10\}\,\si{\radian}$, crossed
with the five seeds for a total of \num{125} (seed, IC) pairs per
controller.

\section{Results}\label{sec:results}

\subsection{Nominal Performance}\label{subsec:nominal}
Table~\ref{tab:nominal} reports the four metrics on the nominal
initial-error recovery task ($x_0=(0.2,0.05,0,0)$). SAC training is
stochastic in its random seed, so the SAC and hybrid entries are
reported as mean $\pm$ standard deviation over the five SAC training
seeds; PID and MPC are deterministic given the plant and initial
condition, so a single value is exact and no spread is reported. SAC achieves the lowest RMSE
($0.0167 \pm 0.0006$\,m), reducing the lateral-error RMSE by
$48\%$ relative to PID and by $41\%$ relative to linear MPC. The
hybrid controller, with the chosen $\lambda = 12$, matches SAC
closely ($0.0171 \pm 0.0006$\,m, a relative gap of $\sim\!2\%$ vs.\
SAC), while remaining within the same actuator envelope as MPC. PID has the largest peak lateral error
($\SI{0.250}{\meter}$); the other three controllers all peak at
$\SI{0.2375}{\meter}$, the unavoidable growth of $\ey$ during the
first sample given the initial conditions and the steering's
one-sample propagation delay through the vehicle dynamics. MPC has the smallest mean
steering effort ($\overline{|\delta|} = \SI{1.55e-3}{\radian}$),
consistent with its explicit input penalty. Across the five seeds,
the standard deviation of every metric for SAC and the hybrid is at
most $6\!\times\!10^{-4}$\,m on RMSE and zero on $\max|\ey|$,
confirming that the result is not driven by a single fortunate
checkpoint.

\begin{table*}[t]
\centering
\small
\setlength{\tabcolsep}{6pt}
\caption{Nominal performance on the initial-error recovery task
($x_0=(0.2,0.05,0,0)$, $T=20$\,s, $\dt=0.05$\,s).
SAC and Hybrid: mean $\pm$ std over five SAC training seeds;
PID and MPC are deterministic.}
\label{tab:nominal}
\begin{tabular}{@{}lcccc@{}}
\toprule
Ctrl. &
RMSE($\ey$)\,[m] &
$\max|\ey|$\,[m] &
$\overline{|\delta|}$\,[rad] &
$\ey(T)$\,[m] \\ \midrule
PID    & 0.0319 & 0.250  & 0.00279 & $\sim\!1\!\times\!10^{-14}$ \\
MPC    & 0.0283 & 0.2375 & 0.00155 & $\sim\!1\!\times\!10^{-37}$ \\
SAC    & $0.0167\!\pm\!0.0006$ & 0.2375 & $0.00302\!\pm\!0.0002$ & $\sim\!4\!\times\!10^{-3}$ \\
Hybrid & $0.0171\!\pm\!0.0006$ & 0.2375 & $0.00290\!\pm\!0.0002$ & $\sim\!4\!\times\!10^{-3}$ \\
\bottomrule
\end{tabular}
\end{table*}

\begin{figure}[t]
\centering
\includegraphics[width=\columnwidth]{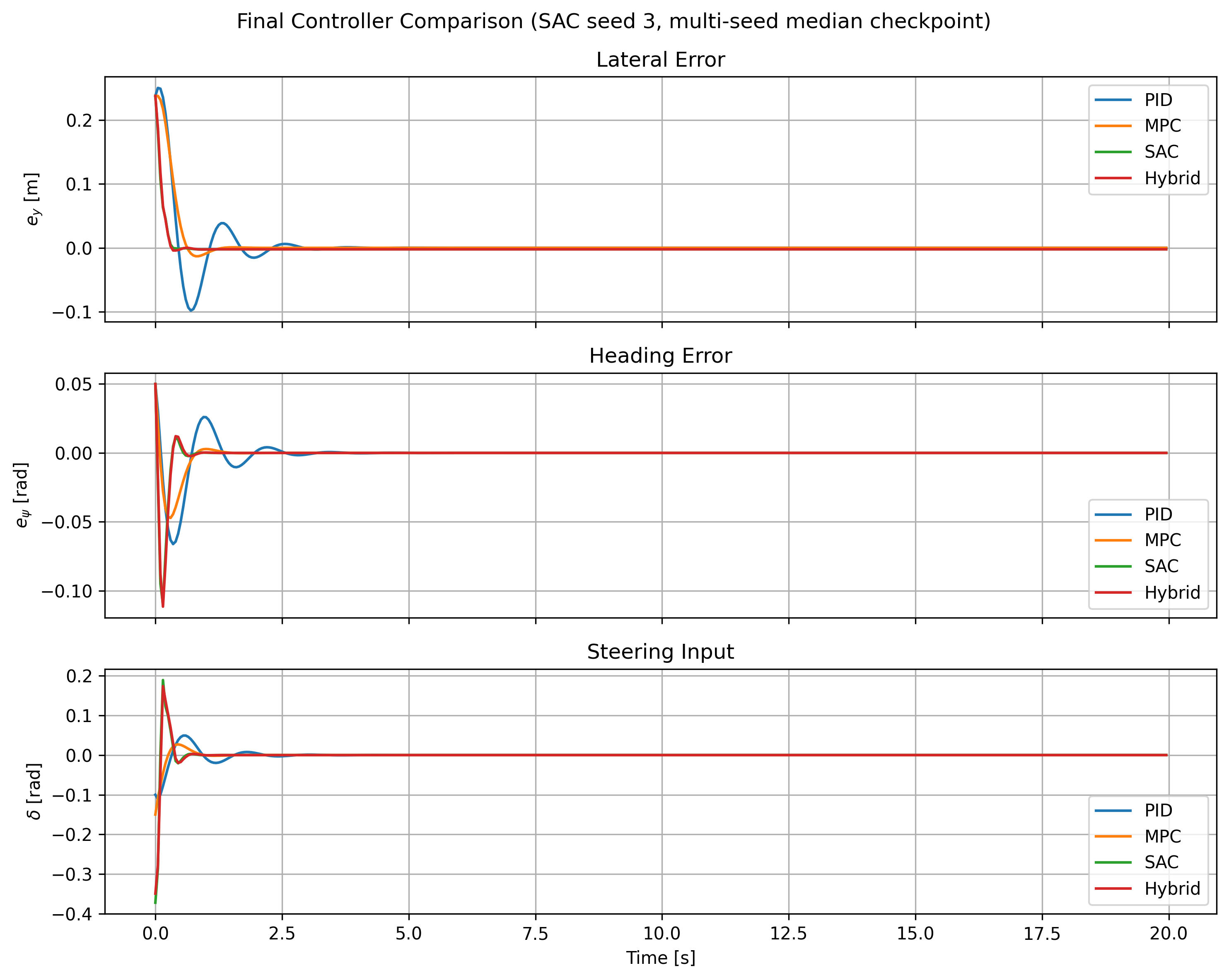}
\caption{Lateral error $\ey$, heading error $\epsiang$, and steering
input $\delta$ for the four controllers on the nominal
initial-error recovery task ($x_0=(0.2,0.05,0,0)$).}
\label{fig:final_comparison}
\end{figure}

The trajectories themselves are shown in
Fig.~\ref{fig:final_comparison}. PID and MPC both drive the lateral
error to numerical zero as $T\to\infty$, while SAC and the hybrid
converge to a small but non-zero residual ($\sim \SI{4}{\milli\meter}$).
This residual is consistent with SAC's stochastic-policy training
objective, which trades a small steady-state bias for lower
transient cost over the training distribution.

\subsection{Hybrid Tuning}
Table~\ref{tab:hybrid} reports the score~\eqref{eq:lamscore} as
$\lambda$ varies, aggregated over the same five SAC training seeds
used in Tables~\ref{tab:nominal} and~\ref{tab:robust}. The composite
score is essentially monotone in $\lambda$: larger $\lambda$ pushes
the blend toward the SAC policy, which has the lowest RMSE on this
task. The best score is achieved at $\lambda = 15$ ($J = 0.1372$);
we use $\lambda = 12$ in the final hybrid because $\lambda = 12$
gives a near-identical score ($J = 0.1373$) while leaving slightly
more weight on the model-based component
($\alpha(12) \approx 0.923$ vs.\ $\alpha(15) \approx 0.938$). The
nominal-case row ($\lambda = 12$) reproduces the
$0.0171\,\si{\meter}$ Hybrid RMSE of Table~\ref{tab:nominal},
confirming consistency between the sweep and the headline
result. All sweeps share the same peak
$\max|\ey| = \SI{0.2375}{\meter}$, indicating that the peak is
governed by the actuator-rate-limited first sample rather than by the
blend.

\begin{table}[t]
\centering
\caption{Hybrid blending sweep over $\lambda$ on the nominal task.}
\label{tab:hybrid}
\small
\setlength{\tabcolsep}{4pt}
\begin{tabular}{@{}lcccc@{}}
\toprule
$\lambda$ &
RMSE($\ey$)\,[m] &
$\max|\ey|$\,[m] &
$\overline{|\delta|}$\,[rad] &
$J$ \\ \midrule
1  & $0.02018\!\pm\!0.00041$ & 0.2375 & 0.00232 & 0.14009 \\
2  & $0.01876\!\pm\!0.00047$ & 0.2375 & 0.00253 & 0.13878 \\
3  & $0.01818\!\pm\!0.00051$ & 0.2375 & 0.00264 & 0.13825 \\
5  & $0.01767\!\pm\!0.00056$ & 0.2375 & 0.00274 & 0.13779 \\
8  & $0.01731\!\pm\!0.00060$ & 0.2375 & 0.00285 & 0.13749 \\
10 & $0.01719\!\pm\!0.00060$ & 0.2375 & 0.00288 & 0.13738 \\
\textbf{12} & $\mathbf{0.01711\!\pm\!0.00060}$ & \textbf{0.2375} & \textbf{0.00290} & \textbf{0.13731} \\
15 & $0.01702\!\pm\!0.00059$ & 0.2375 & 0.00293 & 0.13724 \\
\bottomrule
\end{tabular}
\end{table}

\subsection{Robustness}\label{subsec:robustness}
Table~\ref{tab:robust} reports lateral-error RMSE on the four cases of
Section~\ref{sec:experiments}, aggregated over the five SAC training
seeds. SAC achieves the lowest mean RMSE in every case, reducing it
by $35$-$49\%$ relative to PID and by $32$-$44\%$ relative to MPC.
The hybrid mirrors SAC closely,
preserving the SAC tracking advantage while keeping the actuator
envelope and the deterministic contribution of the modular component
at every step.
Per-seed standard deviation is below $7\!\times\!10^{-4}$\,m for
SAC and the hybrid in every case, confirming that the result is not
driven by an outlier seed.

\begin{table*}[t]
\centering
\small
\setlength{\tabcolsep}{6pt}
\caption{Robustness across the four cases. RMSE($\ey$) in meters,
mean $\pm$ std over five SAC training seeds. PID and MPC are
deterministic.}
\label{tab:robust}
\begin{tabular}{@{}lcccc@{}}
\toprule
Case & PID & MPC & SAC & Hybrid \\ \midrule
A: large $\ey$ ($\ey^{(0)}\!=\!0.4$)             & 0.0522 & 0.0502 & $0.0340\!\pm\!0.0006$ & $0.0347\!\pm\!0.0006$ \\
B: large $\epsiang$ ($\epsiang^{(0)}\!=\!0.10$)  & 0.0446 & 0.0402 & $0.0227\!\pm\!0.0005$ & $0.0234\!\pm\!0.0005$ \\
C: process noise                                 & 0.0318 & 0.0282 & $0.0167\!\pm\!0.0006$ & $0.0171\!\pm\!0.0006$ \\
D: $(C_f,C_r)\!\times\!0.8$                      & 0.0363 & 0.0300 & $0.0179\!\pm\!0.0005$ & $0.0183\!\pm\!0.0005$ \\
\bottomrule
\end{tabular}
\end{table*}

\subsection{Multi-IC Ensemble and Distribution-Shift Stress Test}
\label{subsec:multic}
To stress-test the controllers across the full operating distribution,
we evaluate all four controllers on a $5\!\times\!5$ grid of initial
conditions, $\ey \in \{-0.4,-0.2,0,0.2,0.4\}\,\si{\meter}$ and
$\epsiang \in \{-0.10,-0.05,0,0.05,0.10\}\,\si{\radian}$, crossed
with the five SAC training seeds, for \num{125} (seed, IC) pairs per
controller. We declare a run \emph{divergent} if
$\max_k|\ey(k)| > \SI{1.0}{\meter}$ during the \SI{20}{\second}
episode, and report the divergence rate as well as the lateral-error
RMSE aggregated over the convergent runs only
(Table~\ref{tab:multic}).

PID and MPC are convergent on all \num{125} pairs, with mean
RMSE of $0.0321$\,m and $0.0276$\,m respectively, close to their
nominal-IC values but with larger standard deviation across the IC
grid because it spans a wider distribution of initial errors. SAC and the hybrid are
convergent on \num{122} of \num{125} pairs ($2.4\%$ divergence rate)
and achieve mean RMSE of $0.0176 \pm 0.0112$\,m and
$0.0180 \pm 0.0115$\,m respectively, roughly $36\%$ below MPC across
the entire grid.

The three divergent runs all occur at the double-corner ICs
$(\pm 0.4, \pm 0.10)$ on two of the five training seeds (specifically,
seed 3 at both $(+0.4, +0.10)$ and $(-0.4, -0.10)$, and seed 4 at
$(-0.4, -0.10)$), i.e.\ at the very boundary of the training
distribution, where the
randomized-IC sampler in Section~\ref{sec:sac} has had only
near-zero probability of producing a training example. The hybrid
inherits SAC's failure on exactly these three (seed, IC) pairs:
because the chosen $\alpha(12) \approx 0.92$ leaves only $\sim\!8\%$
of the steering command for the model-based component, the closed-form
blend cannot arrest a saturated SAC action whose sign is already
locked at the actuator limit. This is consistent with the analytical
prediction that the modular contribution is bounded by $(1-\alpha)\,
\delta_{\max} \approx \SI{0.031}{\radian}$ at $\lambda=12$, which
cannot fully counter a saturated end-to-end command in a single
sample. The finding motivates the connectivity-aware $\lambda$
schedule discussed in Section~\ref{sec:discussion}. The idea is to use V2X
signals to detect when the vehicle is operating outside the SAC
training distribution. When such operation is detected, $\lambda$
is reduced, shifting weight in the convex blend toward the MPC
component. With enough MPC weight, the blend can prevent corner
failures rather than merely attenuating their magnitude.

\begin{table*}[t]
\centering
\small
\setlength{\tabcolsep}{6pt}
\caption{Multi-IC ensemble: $5\!\times\!5$ IC grid $\times$ 5 SAC
seeds (\num{125} pairs per controller; PID and MPC are deterministic
and identical across seeds). Divergent: $\max|\ey| >
\SI{1.0}{\meter}$. RMSE and $\overline{\max|\ey|}$ are mean $\pm$
std over the non-divergent runs.}
\label{tab:multic}
\begin{tabular}{@{}lcccc@{}}
\toprule
Ctrl. & Div.\ rate & Conv.\ runs & RMSE($\ey$)\,[m] & $\overline{\max|\ey|}$\,[m] \\ \midrule
PID    & $0/125$\;($0\%$)     & 125 & $0.0321 \pm 0.0156$ & $0.275 \pm 0.131$ \\
MPC    & $0/125$\;($0\%$)     & 125 & $0.0276 \pm 0.0166$ & $0.265 \pm 0.141$ \\
SAC    & $3/125$\;($2.4\%$)   & 122 & $0.0176 \pm 0.0112$ & $0.244 \pm 0.142$ \\
Hybrid & $3/125$\;($2.4\%$)   & 122 & $0.0180 \pm 0.0115$ & $0.244 \pm 0.142$ \\
\bottomrule
\end{tabular}
\end{table*}

\section{Discussion}\label{sec:discussion}
\paragraph{What each side contributes.}
The architectural promise of an end-to-end-plus-modular hybrid is that
each side compensates for the other's weakness. In our experiments
this decomposition is concrete and visible in
Table~\ref{tab:nominal}. The \emph{end-to-end} component (SAC) supplies
the lowest lateral-error RMSE of any controller on the
initial-condition recovery task, including under process noise and
tire-stiffness mismatch (Cases C and D), without any
hand-tuned cost weights; this is the adaptability and
distribution-fitting strength of a learned policy. The \emph{modular}
component (the linear MPC) supplies explicit input and input-rate
constraints, the lowest mean steering effort, an interpretable
per-step output, and a deterministic, model-based contribution at
every step that does not depend on the SAC checkpoint
or on running a neural-network policy at the actuator rate. The
hybrid retains the SAC component's nominal tracking quality while
remaining inside the MPC's actuator envelope and exposing a single,
monotone blending knob $\lambda$ that an integrator or a runtime
monitor can use to move the policy along the modular-end-to-end
axis.

\paragraph{What the hybrid actually is, today.}
The hybrid evaluated in this paper is a closed-form convex blend
between the linear MPC's first-step optimum
$\delta_0^{\star}$ and the SAC policy $\delta_{\mathrm{SAC}}$,
saturated to the MPC actuator envelope. It is \emph{not} the full
constrained-QP predictive safety filter of \eqref{eq:safetyfilter}:
the SAC action does not enter the MPC objective as a soft penalty,
and the rate constraint is enforced only through the MPC's own
prediction (which $\delta_0^{\star}$ inherits) plus the final
saturation, not over the entire blended horizon. The empirical
implication is that the hybrid inherits the per-step interpretability
and constrained, model-aware structure of $\delta_0^{\star}$ and the
asymptotic tracking quality of SAC, but does \emph{not} provide
formal recursive feasibility (a guarantee that the optimization stays
solvable for all future steps) or terminal invariance (a guarantee
that the state remains in a designed safe set). The natural next step is to replace
\eqref{eq:hybridcmd} with the full QP of \eqref{eq:safetyfilter}.

\paragraph{When the hybrid wins.}
On the nominal task and on the four single-axis robustness cases
(Cases A-D), the hybrid tracks stand-alone SAC closely. The value of the hybrid is therefore not
in nominal RMSE but in three operational properties demonstrated by
the multi-IC ensemble (Section~\ref{subsec:multic}): (i) the blending parameter $\lambda$ continuously interpolates
between the modular and end-to-end components, so a system
integrator or runtime monitor can reduce the learned policy's
contribution when distribution shift (states, plant parameters, or
sensor statistics outside the training distribution) is detected; (ii) the modular component
$\delta_0^{\star}$ is the first-step optimum of a constrained QP and
therefore an interpretable, deterministic, model-based contribution
that does not depend on the SAC checkpoint or on running a
neural-network policy at the actuator rate; and (iii) at the chosen
$\alpha(12)\!\approx\!0.92$ the hybrid inherits SAC's $3/125$
($2.4\%$) corner-case divergences, with the analytically clear
caveat that the model-based contribution is bounded by
$(1-\alpha)\,\delta_{\max} \approx \SI{0.031}{\radian}$ and cannot
counter a saturated end-to-end command in a single sample.
Operationally, a fixed-$\lambda$ hybrid is appropriate for broadly
in-distribution operation, and a scheduled $\lambda$ (the
V2X-scheduled extension discussed later in this section) is the
proposed mechanism for handling corner-case failures.

\paragraph{Limitations.}
Four limitations are worth flagging:
\emph{(L1) The hybrid is not a full safety filter.} As above; the
implementation is a closed-form convex blend that omits the predictive
horizon and rate constraint of \eqref{eq:safetyfilter}, and it cannot
prevent failure on out-of-distribution inputs (only cap its
magnitude, as observed in Section~\ref{subsec:multic}). \emph{(L2)
Linearized lateral bicycle, not a full nonlinear vehicle.} Tire
saturation, longitudinal coupling, and road-friction effects are not
modeled. \emph{(L3) Constant longitudinal speed.} Combined
longitudinal-lateral control is left to future work. \emph{(L4)
Open-loop sim only.} No CarSim/CARLA validation, no on-vehicle test.

\paragraph{Path forward.}
The most direct upgrades are: replace the closed-form blend with the
constrained QP \eqref{eq:safetyfilter}; substitute the linear
bicycle for a higher-fidelity model with tire-curve saturation;
extend the action to longitudinal acceleration; and validate the
trained SAC and hybrid in a co-simulator before any real-vehicle work.

\paragraph{Connectivity-aware extension.}
The blending coefficient $\lambda$ in \eqref{eq:hybridcmd} is more than
a tuning knob: it is an explicit, low-bandwidth interface between the
controller and any source of information about the current
operational design domain. Concretely, a runtime monitor can compute
$\lambda$ from three classes of distribution-shift signals:
(i)~direct state monitoring of $|\ey|$ and $|\epsiang|$ against the
training-distribution bounds, which captures state-distribution
shifts; (ii)~vehicle-to-everything (V2X) messages such as
road-friction estimates from preceding vehicles, infrastructure
weather and surface reports, and local-map confidence, which carry
plant-parameter shifts (e.g.\ effective tire stiffness under wet,
icy, or snow conditions) before they appear in the state; and
(iii)~onboard model-residual monitoring (e.g.\ innovations from a
Kalman filter comparing predicted versus measured state evolution),
which catches both plant-parameter and sensor-statistics shifts.

A natural extension of the architecture presented here is to
schedule $\lambda(\text{V2X})$ at runtime: increase $\lambda$
(toward the learned policy) when these signals confirm
in-distribution operation, and decrease it (toward the modular
controller) when any source reports a degraded scenario. The fact
that $\lambda$ is a single, smooth, monotone parameter, rather than
a discrete mode switch, makes this scheduling stable and amenable
to formal analysis. A V2X-scheduled $\lambda$ that dials $\alpha$
down to $\alpha(1)=0.5$
(equal authority for both components) or further to $\alpha(0.1)
\approx 0.09$ (essentially MPC) would expand the modular contribution
to as much as $\delta_{\max}=\SI{0.4}{\radian}$ in a single sample,
which is sufficient to convert these residual corner failures into
bounded tracking errors of the same order as the pure MPC, at the
cost of slightly worse nominal RMSE. This is precisely the
trade-off the architecture is designed to expose, and we see it as
the most direct way to take the architecture from an
automated-vehicle setting into a connected-and-automated-vehicle
setting.

\section{Conclusion}\label{sec:conclusion}
We presented a hybrid end-to-end and modular control architecture for
safe vehicle lateral control, instantiated by combining a Soft
Actor-Critic policy with a linear, constrained model predictive
controller, and benchmarked it against PID, MPC, and stand-alone SAC
on a linearized lateral bicycle plant. Aggregated over multiple
independent SAC training seeds, the end-to-end component supplies
the lowest nominal lateral-error RMSE; the modular component
supplies explicit input and input-rate constraints, the lowest mean
steering effort, and a deterministic contribution at every step.
The hybrid retains the tracking quality of stand-alone SAC while
exposing a single, monotone blending coefficient $\lambda$ that
interpolates between the constrained linear MPC and the learned
policy, inside the MPC's actuator envelope. A multi-initial-condition
ensemble across the IC grid crossed with the SAC seeds confirms
that the SAC and hybrid controllers retain their tracking advantage
relative to PID and MPC on the vast majority of (seed, IC) pairs;
on a small minority at the corners of the training distribution,
the closed-form blend exhibits the same divergences as stand-alone
SAC, motivating the connectivity-aware schedule introduced in
Section~\ref{sec:discussion}. The implementation evaluated here is
a closed-form convex blend, while the architecture is designed to
support the full constrained-QP predictive safety filter of
\eqref{eq:safetyfilter}. The two most direct next steps are this
full-QP version and a connectivity-aware schedule
$\lambda(\text{V2X})$. A companion paper is in preparation that
implements the full constrained-QP predictive safety filter of
Eq.~\eqref{eq:safetyfilter} with a terminal invariant set for
formal recursive feasibility, empirically validates the
V2X-scheduled $\lambda$ extension, and evaluates the architecture
on a nonlinear single-track vehicle model.



\end{document}